%% file: main.tex
\documentclass[reprint,superscriptaddress,amsmath,amssymb,prb]{revtex4-1}
\usepackage[dvipdfmx]{graphicx}
\usepackage{dcolumn}
\usepackage{bm}
\usepackage{color}
\usepackage{amssymb}
\usepackage{comment}

\begin{document}

\title{Subcycle control of valley-selective excitation via dynamical Franz-Keldysh effect in WSe$_2$ monolayer}

\author{Shunsuke Yamada}
\affiliation{Kansai Photon Science Institute, National Institutes for Quantum Science
and Technology (QST), Kyoto 619-0215, Japan}
\author{Kazuhiro Yabana}
\affiliation{Center for Computational Sciences, University of Tsukuba, Tsukuba 305-8577, Japan}
\author{Tomohito Otobe}
\affiliation{Kansai Photon Science Institute, National Institutes for Quantum Science
and Technology (QST), Kyoto 619-0215, Japan}

\date{\today}

\begin{abstract}
This study performed first-principles calculations based on the time-dependent density functional theory to  control the valley degree of freedom relating to the dynamical Franz-Keldysh effect (DFKE) in a monolayer of transition metal dichalcogenide.
By mimicking the attosecond transient absorption spectroscopy, we performed numerical pump-probe experiments to observe DFKE around the $K$ or $K'$ valley in WSe$_2$ monolayer with a linearly-polarized pump field and a circularly-polarized probe pulse.
We found that the circularly-polarized probe pulse with a given helicity can selectively observe the transient conductivity modulated by DFKE in each valley.
The transient conductivity and excitation probability around each valley oscillate with the pump field frequency $\Omega$.
The phases of the $\Omega$ oscillation for the $K$ and $K'$ valleys are opposite to each other.
Furthermore, the pump-driven DFKE alters the absorption rate of WSe$_2$ monolayer and yields the valley-dependent $\Omega$ oscillation of the electron excitation induced by the pump plus probe field.
With a simplified two-band model, we identified the $\Omega$ oscillation of the off-diagonal conductivity caused by the band asymmetry around the valleys as the physical mechanism responsible for the valley-selective DFKE.
\end{abstract}
\maketitle

\section{Introduction}

Valleytronics has emerged as an active area of research for controlling the valley degree of freedom in  certain semiconductors that contain multiple valleys in their electronic band structure\cite{Behnia2012,Nebel2013}. 
The fundamental goal in valleytronics research is to lift the valley degeneracy and create valley polarization to store and manipulate the bits of information. 
In general, two-dimensional (2D) hexagonal lattice materials possess two valleys at the $K$ and $K'$ points in the 1st Brillouin zone, and constitute a promising direction in  valleytronics\cite{Xiao2007,Yao2008,Xu2014,Schaibley2016,Vitale2018}. 

Recently, ultrafast valley control techniques in 2D materials using optical pulses have been actively investigated for potential applications in ultrafast signal processing\cite{Zeng2012,Mak2012,Yuan2014,Langer2018,OliaeiMotlagh2018,OliaeiMotlagh2019,Nematollahi2020,Jimenez-Galan2020,Jimenez-Galan2021,Kumar2021,Mrudul2021}.
In particular, transition metal dichalcogenide (TMDC) monolayers are attractive for practical applications owing to their broken inversion symmetry and strong spin-orbit coupling (SOC)\citep{Xu2014}.
SOC lifts the spin degeneracy in the $K$ and $K'$ valleys and yields the opposite spin angular momenta owing to the time-reversal symmetry. 
The spin-valley locking and the interplay of the valleys yield the valley-dependent optical selection rules\cite{Xiao2012,Xu2014}.
Specifically, the interband transition at the $K$ ($K'$) valley is exclusively coupled to the left (right) circularly polarized light resonant with the bandgap.


In addition to the ultrafast optical valley control,  there is another active area of 
research for developing ultrafast signal processing.
The dynamical Franz-Keldysh effect (DFKE) is a phenomenon that occurs in dielectrics under the irradiation of an alternating electric field with an off-resonant frequency\cite{Yacoby1968,Jauho1996,Nordstrom1998,Srivastava2004,Mizumoto2006,Ghimire2011,Chin2000}. 
Upon applying a strong alternating electric field to a crystalline dielectric, it causes an intraband motion of charges and induces a transient change in the optical response, which oscillates at a frequency of multiples of the field frequency. 
Unlike resonant processes, DFKE is an ultrafast nonresonant process that does not excite real carriers. 
Recent advancements in experimental techniques such as the attosecond transient absorption spectroscopy (ATAS)\cite{Goulielmakis2010,Wang2010,Holler2011,Gaarde2011,Beck2015,Wu2016} have enabled the observation of DFKE in the petahertz regime (femtosecond time scales)\cite{Schultze2013,Schultze2014,Mashiko2016,Lucchini2016,Moulet2017,Schlaepfer2018,Lucchini2020,Lucchini2021,Buades2021}.
Therefore, DFKE is expected to be a potential candidate for ultrafast optical switching in future petahertz signal processing.
In recent years, subcycle DFKE in 2D materials\cite{Sato2018,Cistaro2021,Dong2022} as well as solids\cite{Otobe2016,Otobe2016-2,Otobe2017} has been theoretically investigated.

This study explored the potential for combining these two research areas by investigating the applicability of DFKE for ultrafast valley switching in TMDC monolayers.
We performed first-principles calculations based on the time-dependent density functional theory (TDDFT)\cite{Runge1984} to explore the possibility of  valley-selective ultrafast optical switching using DFKE in a laser-irradiated TMDC monolayer. 
We focused on WSe$_2$ monolayer and considered a control technique of the phase of the DFKE oscillation depending on the valley degree of freedom.
To mimic measurements of ATAS, we conducted numerical pump-probe experiments similar to that in previous theoretical studies\cite{Otobe2016,Otobe2016-2,Otobe2017,Sato2018}.
A semi-infinite monochromatic light with an off-resonant frequency was employed as the pump field and a left (right) circular-polarized ultrashort pulse was used as the probe field to observe  DFKE around the $K$ ($K'$) valley.
From the band asymmetry around the $K$ and $K'$ valleys, it may be expected that the phase of the DFKE oscillation near the bandgap is inverted depending on the left or right circular polarization of the probe pulse.
Furthermore, DFKE may alter the absorption rate of WSe$_2$ monolayer and yield the valley-dependent oscillation of the electron excitation induced by the pump plus probe field.
Using a simplified two-band model, we discussed the physical mechanisms behind the valley-selective phase inversion of the DFKE oscillation.

The remainder of this paper is organized as follows: 
The theoretical formalism of real-time TDDFT calculations along with the simplified two-band model are presented in Sec.~\ref{sec:theory}.
The results derived from the real-time first-principles and two-band model calculations are presented and discussed in Sec.~\ref{sec:results}. 
Lastly, the conclusions of this study are summarized in Sec.~\ref{sec:conclusion}.

\section{Theoretical formalism \label{sec:theory}}

\begin{figure}
    \includegraphics[keepaspectratio,width=\columnwidth]{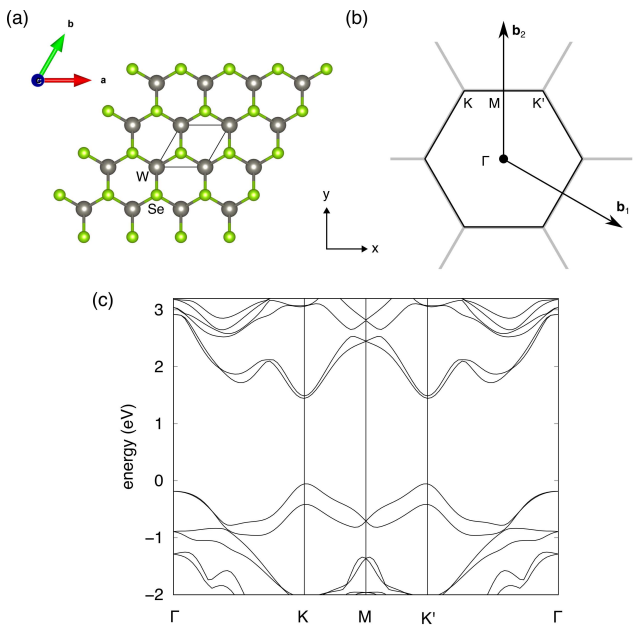}
    \caption{\label{fig:band} 
    (a) Atomic configuration,
    (b) Brillouin zone, and
    (c) band structure of WSe$_2$ monolayer.
    }
\end{figure}

In this section, we present the theoretical framework aiming  to describe the valley-selective DFKE in a TMDC monolayer driven by an intense light field.
We focus on WSe$_2$ monolayer and consider its transient optical properties in time-dependent calculations mimicking pump-probe measurements.

First, we look at the electronic structure of the WSe$_2$ monolayer to devise a strategy.
The atomic configuration, Brillouin zone, and band structure of the WSe$_2$ monolayer are presented in Fig.~\ref{fig:band}(a), (b), and (c), respectively.
The band structure in Fig.~\ref{fig:band}(c) is depicted along the $x$ axis, i.e., zigzag direction in real space.
The high-symmetry points $K$ and $K'$ correspond to the direct bandgaps, and all bands are split by the intrinsic spin-orbit coupling, except at the time-reversal invariant $\Gamma$ and $M$ points.
Although the band structure near the  $K$ ($K'$) point is asymmetric with respect to the $K$ ($K'$)  point, the entire band structure is reflection symmetric with respect to the $M$ point in the $x$-axis.
Moreover, a normally incident circularly-polarized light  exclusively excites electrons near the $K$ or $K'$  valleys depending on the left or right circular polarization~\cite{Xiao2012,Xu2014}.

Based on this perspective, the valley-selective DFKE can be achieved by the procedure stated as follows.
First, the WSe$_2$ monolayer is irradiated with a monochromatic $x$-polarized light with a frequency substantially lower than the bandgap.
The light field induces an oscillation of the optical property change in the temporal domain because of DFKE around the $K$ and $K'$ valleys.
According to  the reflection symmetry of the band structure in the $x$-axis, the phases of the DFKE oscillation in the $K$ and $K'$ valleys may be opposite to each other.
Upon further applying a normally incident  probe pulse with  left or right circular polarization to the system, we may obtain the transient optical response corresponding to the $K$ or $K'$ valley.
As the DFKE oscillations around the $K$ and $K'$ valleys may exhibit phases opposite to each other, the phase of the transient optical response is inverted depending on the left or right polarization of the probe pulse.
This phase flip corresponds to the selection of the $K$ and $K'$ valleys, and thus, the valley-selective DFKE may be realized by the $x$-polarized pump and circularly-polarized probe light fields.

In the present study, we perform numerical pump-probe experiments for the valley-selective DFKE in the WSe$_2$ monolayer  using two computational methods: a first-principles method based on TDDFT and a simple two-band model described below.
Based on time profiles of the  current density induced by the pump and probe electric fields, we extract the transient optical properties including the DFKE oscillation.
To this end, we discuss the optical conductivity tensor of the TMDC monolayers and develop a procedure for extracting the transient optical properties from the calculation results.

\subsection{TDDFT}

To realistically simulate pump-probe measurements for DFKE, we apply a TDDFT formalism for electron dynamics in  presence of an electric field\cite{Bertsch2000,Otobe2008}.
We consider electron motion in a TMDC monolayer under irradiation of the electric field ${\bf E}(t)=-(1/c)d{\bf A}(t)/dt$  in the dipole approximation.
The time-dependent Kohn-Sham (TDKS) equation for the Bloch orbital
$u_{b,{\bf k}}({\bf r},t)$ (a two-component spinor, where $b$ denotes the band index and ${\bf k}$ indicates the 2D crystal momentum of the
2D material) is described as follows:
\begin{equation}
\begin{split}i\hbar\frac{\partial}{\partial t}u_{b,{\bf k}}({\bf r},t)=\Big[\frac{1}{2m}{\left(-i\hbar\nabla+\hbar{\bf k}+\frac{e}{c}{\bf A}(t)\right)}^{2}\\
-e\varphi({\bf r},t)+\hat{v}_{{\rm NL}}^{{{\bf k}+\frac{e}{\hbar c}{\bf A}(t)}}+{v}_{{\rm xc}}({\bf r},t)\Big]u_{b,{\bf k}}({\bf r},t),
\end{split}
\label{eq:tdks}
\end{equation}
where the scalar potential $\varphi({\bf r},t)$ includes the Hartree
potential from the electrons and the local component of the ionic pseudopotentials and we have defined  $\hat{v}_{{\rm NL}}^{{\bf k}}\equiv e^{-i{\bf k}\cdot{\bf r}}\hat{v}_{{\rm NL}}e^{i{\bf k}\cdot{\bf r}}$. 
Here, $\hat{v}_{{\rm NL}}$ and ${v}_{{\rm xc}}({\bf r},t)$ represent the
nonlocal component of the ionic pseudopotentials and exchange-correlation
potential, respectively. 
The spin-orbit coupling is incorporated through the $j$-dependent nonlocal potential $\hat{v}_{{\rm NL}}$ \cite{Theurich2001}, and the Bloch orbitals $u_{b,{\bf k}}({\bf r},t)$
are defined in a box containing the unit cell of the TMDC monolayer sandwiched by vacuum regions. 

The 2D current density (electric current per unit area) ${\bf J}(t)$ is derived from the Bloch orbitals as follows:
\begin{equation}
\begin{split}{\bf J}(t)=-\frac{e}{m}\int dz\int_{\Omega}\frac{dxdy}{N_k \Omega}\sum_{b,{\bf k}}^{{\rm occ}}u_{b,{\bf k}}^{\dagger}({\bf r},t)\\
\times\left[-i\hbar\nabla+\hbar{\bf k}+\frac{e}{c}{\bf A}(t)+\frac{m}{i\hbar}\left[{\bf r},\hat{v}_{{\rm NL}}^{{{\bf k}+\frac{e}{\hbar c}{\bf A}(t)}}\right]\right]u_{b,{\bf k}}({\bf r},t),
\end{split}
\end{equation}
where $\Omega$ denotes the area of the 2D unit cell and $N_k$ denotes the number of $k$-points.
The sum is taken over the occupied bands in the ground state.
The excited electron population is defined as,
\begin{equation}
    \rho_{\bf k}(t) = \sum_{c,v} \left| 
    \int_{\Omega} d^3 r \,
    u_{v, {\bf k}}^{\dagger}({\bf r},t) \,
    u_{c, {\bf k} + \frac{e}{\hbar c} {\bf A}(t) }^{\rm GS}({\bf r})
    \right|^2,
\end{equation}
where $v$ and $c$ denote the indices for the valence and conduction bands, respectively, and $u_{b, {\bf k} }^{\rm GS}({\bf r}) = u_{b, {\bf k} }({\bf r},t=0) $ indicates the Bloch orbital in the ground state.

The number of excited electrons around the $K$ point is defined as follows:
\begin{equation}
    n_{{\rm ex},K}(t) = \frac{1}{N_k}\sum_{|{\bf k}-{\bf k}_K|<k_{\rm rad}} \rho_{\bf k}(t),
    \label{eq:nex_tddft}
\end{equation}
where the sampling points are considered within a radius of $k_{\rm rad}=$ 0.15 a.u. and ${\bf k}_K$ represents the $k$-vector corresponding to the $K$ point.
The definition for  the $K'$ point is the same as Eq.~(\ref{eq:nex_tddft}) but with  ${\bf k}_{K'}$.

In this paper, the first-principles TDDFT calculations are performed using SALMON code\cite{SALMON_web,Noda2019}.
The calculation conditions are almost the same as those employed in Refs.~\onlinecite{Hashmi2022a} and \onlinecite{Hashmi2022}.
The lattice constant of WSe$_{2}$ monolayer is set to $a = b =$ 3.32 {\AA}. 
The adiabatic local spin density approximation with Perdew-Zunger functional~\cite{Perdew1981} is used for the exchange correlation.
A slab approximation is used for the $z$ axis with a distance of 20 {\AA} between the atomic monolayers.
Although the dynamics of the 24 valence electrons are treated explicitly, the effects of the core electrons are considered through norm-conserving pseudopotentials from the OpenMX library~\cite{Ozaki2003,Ozaki2004,VPS_web}. 
The spatial grid sizes and k-points are optimized according to the converging results. 
The determined parameter of the grid size is 0.21 {\AA}, and the optimized k-mesh is 16 $\mathrm{\times}$ 16 in the 2D Brillouin zone. 

\subsection{Two-band model \label{sec:model}}

Although the TDDFT-based first-principles calculations  provide realistic and reliable descriptions for our problem, the TDDFT results cannot be meaningfully interpreted  without simplifying  the physical process.
To obtain insights into the physical mechanisms determining the TDDFT results, we  perform model calculations using a minimal band model~\cite{Xiao2012,Rostami2013,Kormanyos2013,Liu2013,Berkelbach2015,Kormanyos2015,Hashmi2022a}.
The model Hamiltonian including the second-order coupling for the
low-energy physics prevailing in the $K$ ($K'$) point is described as follows:
\begin{eqnarray}
    H^{\tau ,s}[{\bf k}]
=
\begin{bmatrix}
\frac{\Delta}{2} & a\tilde{t} (\tau k_x -i k_y) 
\\
a\tilde{t} (\tau k_x +i k_y) & -\frac{\Delta}{2}
\end{bmatrix}
\nonumber \\
+ a^2 
\begin{bmatrix}
\gamma_1 k^2 & \gamma_3 (\tau k_x +i k_y)^2 
\\
\gamma_3 (\tau k_x -i k_y)^2 & \gamma_2 k^2
\end{bmatrix}
\nonumber \\
+
\begin{bmatrix}
0 & 0
\\
0 & \tau s \lambda
\end{bmatrix},
\label{eq:model}
\end{eqnarray}
where  $\tau=+1\,(-1)$ denotes the pseudo-spin index and we have redefined the $k$-vector as  ${\bf k}- {\bf k}_K$ (${\bf k}- {\bf k}_{K'}$) $\longrightarrow {\bf k}$.
The first and second terms denote the massive Dirac Hamiltonian and its second-order correction, respectively. 
The third term represents the spin-orbit coupling Hamiltonian, and $s=\pm1$ indicates the spin index.
The parameters $a$, $\Delta$, $\tilde{t}$, and $\lambda$ represent the lattice constant, bandgap, hopping parameter, and spin-orbit splitting of the valence
band, respectively.
The parameters $\gamma_{1}$ and $\gamma_{2}$ represent the breaking
of the electron-hole symmetry, and the parameter $\gamma_{3}$ is responsible
for the band asymmetry. 
These parameters are determined by fitting
the  band structure calculated using SALMON\cite{Hashmi2022a}. 

The electron dynamics in the presence of the electric field ${\bf E}(t)=-(1/c)d{\bf A}(t)/dt$ can be described using 
\begin{equation}
    i\hbar \frac{d}{dt}\psi^{\tau,s}_{\bf k}(t) = H^{\tau,s}\left[{\bf k}+\frac{e}{\hbar c}{\bf A}(t)\right] \psi^{\tau,s}_{\bf k}(t),
    \label{eq:td_model}
\end{equation}
where $\psi^{\tau,s}_{\bf k}(t) = (\psi^{{\tau s}{\bf k}}_{1}(t),\psi^{{\tau s}{\bf k}}_{2}(t))^{T}$ denotes the time-dependent wavefunction.
The initial value of the wavefunction is set to the valence band wavefunction at the ground state.
The 2D current density is expressed as follows:
\begin{equation}
	{\bf J}(t)= 
    -\frac{c}{N_{k}}\sum_{\tau,s,{\bf k}}
    \langle \psi_{\bf k}^{\tau,s}(t) |
    \frac{\partial H^{\tau,s}\left[{\bf k}+({e}/{\hbar c}){\bf A}(t)\right] }{\partial {\bf A}(t)}
    |\psi_{\bf k}^{\tau,s}(t)\rangle,
\end{equation}
where the $k$-point sampling is  within the radius of $k_{\rm rad}=$ 0.15 a.u. around  ${\bf k}=0$  and $N_{k}$ denotes the number of the sampling points. 
The number of excited electrons or excitation probability from the valence band to the conduction band is derived as 
\begin{equation}
n_{\rm ex}^{\tau}(t)=\frac{1}{N_{k}}\sum_{s,{\bf k}}\left|\langle\phi_{{\rm c},{\bf k}+\frac{e}{\hbar c} {\bf A}(t)}^{\tau,s}|\psi_{\bf k}^{\tau,s}(t)\rangle\right|^{2},
\label{eq:nex_model}
\end{equation}
where $\phi_{{\rm c},{\bf k}}^{\tau,s}$ denotes the conduction band wavefunction at the ground state.

\subsection{Transient optical properties with pump-probe calculations}

Herein, we review a procedure for calculating the transient optical properties of a laser-irradiated material by numerical pump-probe experiments\cite{Otobe2016}.
To observe DFKE, the optical absorbance of a material is measured in the ATAS experiments, and it is proportional to the real part of the optical conductivity.
In the absence of an intense laser field, the optical conductivity tensor $\sigma_{\alpha\beta}(\omega)$ of a 2D material is defined by the following constitutive relation:
\begin{equation}
J_{\alpha}(\omega)=\sum_{\beta}\sigma_{\alpha\beta}(\omega)E_{\beta} (\omega),
\label{eq:constitutive}
\end{equation}
where ${\bf E}(\omega)$ and ${\bf J}(\omega)$ denote the Fourier transforms of the electric field and 2D current density, respectively, in the linear regime.
This optical conductivity in the ground state can be evaluated by a linear response method of TDDFT for 2D materials\cite{Yamada2018}.

To investigate the transient change in the optical properties of the laser-irradiated material, we consider electron dynamics under pump ${\bf E}^{\rm pump}(t)$ and probe ${\bf E}^{\rm probe}(t)$ electric fields. 
As the probe field ${\bf E}^{\rm probe}(t)$, we apply an  ultrashort pulse of sufficiently short duration to observe the pump-driven conductivity change with the subcycle temporal resolution.
The timing of the probe pulse relative to the pump field is specified by the delay time $T_{\rm delay}$.
By solving the time-evolution equation (\ref{eq:tdks}) or (\ref{eq:td_model}) considering both the pump and probe fields, we obtain the current density induced by the electric fields,
which is hereinafter referred to as  pump-probe current density ${\bf J}^{\rm pump+probe}(t)$.
Moreover, we can evaluate the current density under only the pump field, denoted as the pump current density ${\bf J}^{\rm pump}(t)$.
To extract the current density induced by the probe field in the presence of the pump field, the probe current density ${\bf J}^{\rm probe}(t)$ is defined as follows:
\begin{equation}
    {\bf J}^{\rm probe}(t)={\bf J}^{\rm pump+probe}(t)- {\bf J}^{\rm pump}(t).
    \label{eq:probe_current}
\end{equation}
The transient conductivity $\sigma_{\alpha\beta}(\omega, T_{\rm delay})$ in the presence of the pump field can be related as follows:
\begin{equation}
J_{\alpha}^{\rm probe}(\omega)=
\sum_{\beta}
\sigma_{\alpha\beta}(\omega, T_{\rm delay})
  E^{\rm probe}_{\beta}(\omega),
  \label{eq:sigma_probe_tensor}
\end{equation}
where ${\bf J}^{\rm probe}(\omega)$ and ${\bf E}^{\rm probe}(\omega)$ represent the Fourier transforms of the probe current density and probe electric fields, respectively; ${\bf J}^{\rm probe}(\omega)$ and ${\bf E}^{\rm probe}(\omega)$ implicitly depend on  $T_{\rm delay}$.
 In the absence of the pump field, $\sigma_{\alpha\beta}(\omega, T_{\rm delay})$ should be equal to the ordinary conductivity tensor $\sigma_{\alpha\beta}(\omega)$ in Eq.~(\ref{eq:constitutive}).

In this study, we use a circularly-polarized light pulse as the probe field.
We consider the probe electric field ${\bf E}^{{\rm probe}\pm}(t)$ and probe current density ${\bf J}^{{\rm probe}\pm}(t)$ in the TMDC monolayer with the circular ($\pm$) polarization.
Using the unit vectors for the left ($+$) and right ($-$) circular polarization directions, $\hat{\bf e}_{\pm}=(\hat{\bf x}\pm i\hat{\bf y})/\sqrt{2}$, the ``circular" components of a vector or tensor can be derived.
The circular components of the transient conductivity can be expressed as follows:
\begin{eqnarray}
    \sigma_{\pm\pm}(\omega,T_{\rm delay}) &=& \frac{J^{{\rm probe}\pm}_{\pm} (\omega)}{ E^{{\rm probe}\pm}_{\pm} (\omega)}
    \nonumber\\
    &=& \frac {[ J^{{\rm probe}\pm}_{x} (\omega)\mp i J^{{\rm probe}\pm}_{y} (\omega)]/\sqrt{2}}{ [E^{{\rm probe}\pm}_{x} (\omega) \mp i E^{{\rm probe}\pm}_{y} (\omega)]/\sqrt{2}}
    \nonumber\\
    &=&
    \frac{1}{2} \left[
    \sigma_{xx}(\omega,T_{\rm delay}) \pm i \sigma_{xy}(\omega,T_{\rm delay}) \right.
    \nonumber\\
    && \left. \mp i \sigma_{yx}(\omega,T_{\rm delay}) + \sigma_{yy}(\omega,T_{\rm delay})
    \right],
    \nonumber\\
    \label{eq:sigma_circular_probe}
\end{eqnarray}
where $\omega > 0$ and we have used ${E}^{{\rm probe}\pm}_y(\omega)= \pm i{E}^{{\rm probe}\pm}_x(\omega)$ and Eq.~(\ref{eq:sigma_probe_tensor}).
Without the pump field, $\sigma_{\pm\pm}(\omega,T_{\rm delay})$ should be equivalent to the ground state conductivity $\sigma_{xx}(\omega)=\sigma_{yy}(\omega)$, where the off-diagonal elements are zero in the ground state\cite{Have2019,Caruso2022}.
 
\subsection{Consideration for experiments\label{sec:exp}}
For future experimental measurements, we consider the observable quantities in the proposed framework.
The transient absorbance of a target material is measured in the ATAS experiments.
In particular, the absorbed energy of the probe pulse, or the work done by the probe pulse, at each time delay can be derived as 
\begin{align}
    & W^{{\rm probe}\pm}(T_{\rm delay}) 
    =
    \int dt \, {\bf J}^{{\rm probe}\pm}(t) \cdot {\bf E}^{{\rm probe}\pm}(t) 
    \nonumber\\
    & \quad=  2\int_0^{\infty} \frac{d\omega}{\pi}  
    \left| {E}^{{\rm probe}\pm}_{x}(\omega) \right|^2 \,
    {\rm Re}\, \sigma_{\pm\pm}(\omega,T_{\rm delay}).
    \label{eq:work}
\end{align}
Based on this relation, a transient absorption spectrum measured by the ATAS experiments is proportional to ${\rm Re}\,\sigma_{\pm\pm}(\omega,T_{\rm delay})$.
The difference in the absorbed energy between the left and right polarization is stated as follows:
\begin{align}
    & W^{{\rm probe}+}(T_{\rm delay}) - W^{{\rm probe}-}(T_{\rm delay}) 
    \nonumber\\
    & \quad = 2\int_0^{\infty} \frac{d\omega}{\pi}  
    \left| {E}^{{\rm probe}+}_{x}(\omega) \right|^2 
    \nonumber\\
    & \qquad \quad \times {\rm Re} \left[
     \sigma_{++}(\omega,T_{\rm delay}) - \sigma_{--}(\omega,T_{\rm delay})
    \right] 
    \nonumber\\
    & \quad = 2\int_0^{\infty} \frac{d\omega}{\pi}  
    \left| {E}^{{\rm probe}+}_{x}(\omega) \right|^2 
    \nonumber\\
    & \qquad \quad \times {\rm Im} \left[
     \sigma_{yx}(\omega,T_{\rm delay}) - \sigma_{xy}(\omega,T_{\rm delay})
    \right].
    \label{eq:work_diff}
\end{align}
As this value should be zero in the absence of the pump field\cite{Caruso2022}, it can be regarded as  a characteristic observable quantity in the pump-probe system.
Note that a first-principles study has been reported for the circular dichroism in transient absorption spectra for a 2D topological insulator with a circularly-polarized pump pulse\cite{Neufeld2023}.

Furthermore, the valley polarized excitation induced by both the fields in the pump-probe system can be directly measured using the free-carrier valley Hall effect\cite{Mak2014,Vitale2018,Liu2019}.
DFKE induced by the pump field may alter the absorption rate of the TMDC monolayer and yield a valley-dependent oscillation of the excited electron population relating to the probe pulse.
Considering the excited electron population instead of the conductivity change, the proposed system is more suitably defined as a ``double-pump experiment'' rather than a ``pump-probe experiment,'' because the excitation by the probe pulse may be relatively intense compared to that by the pump field.
This is because the pump (1st pump, to be precise) pulse is a strong but off-resonant $x$-polarized field, whereas the probe  (2nd pump, to be precise) pulse is a weak but on-resonant circularly-polarized field.
At the end of Sec.~\ref{sec:tddft_results}, we discuss the valley polarized excitation induced by the ``double-pump'' pulses.

\subsection{Pulse settings}

In the present calculations, we focus on DFKE around the bandgap in the WSe$_{2}$ monolayer.
As the pump field, we use a semi-infinite monochromatic light with a frequency below the bandgap.
To observe the DFKE oscillation of the transient conductivity, we apply an ultrashort circularly-polarized pulse as the probe field.
The frequency range of the probe pulse should be well-separated from that of the pump field to obtain clear DFKE signals\cite{Yamada2020}.

The vector potential for the $x$-polarized pump field is expressed as
\begin{equation}
    {\bf A}^{\rm pump}(t) = - \frac{c E^{\rm pump}_{\rm max}}{\Omega} f(t) \sin \left\{ \Omega \left( t
  - T^{\rm pump} \right) \right\} \hat{\bf x},
  \label{eq:pump}
\end{equation}
where  $\hbar \Omega=0.3$~eV and $T^{\rm pump}=25$~fs.  
$E^{\rm pump}_{\rm max}$ is set to provide the peak intensity of $I^{\rm pump}=5\times 10^{10}$~W/cm$^2$.
$f(t)$ refers to an envelope function defined as follows:
\begin{equation}
    f(t)=
    \begin{cases}
    0, & t<0, \\
    \sin^6 \left( \frac{\pi t}{2T^{\rm pump}}\right), & 0<t<T^{\rm pump}, \\
    1, & T^{\rm pump}<t.
    \end{cases}
\end{equation}
The vector potential for the circularly-polarized probe pulse is expressed as
\begin{eqnarray}
{\bf A}^{\rm probe \pm}(t) 
= - \frac{c E^{\rm probe}_{\rm max}}{\omega} \,  
\cos^6 \left\{ \frac{\pi }{T^{\rm probe}} (t-T_{\rm delay}) \right\}
\nonumber \\
\times \left[ 
  \hat{\bf x} \sin \left\{ \omega \left( t
  - T_{\rm delay} \right) \right\}
  \pm \hat{\bf y} \cos \left\{ \omega \left( t
  - T_{\rm delay} \right) \right\}
\right]
, \nonumber \\ \left( -{T^{\rm probe}}/{2} < t-T_{\rm delay} < {T^{\rm probe}}/{2} \right),
\label{eq:probe}
\end{eqnarray}
where $\hbar \omega=2$~eV and $T^{\rm probe}=10$~fs. 
$E^{\rm probe}_{\rm max}$ is set to provide the peak intensity of $I^{\rm probe}=10^{10}$~W/cm$^2$.

The Fourier transform of the probe current density is defined as follows:
\begin{equation}
 {\bf J}^{\rm probe}(\omega)
 =\int^{T_{\rm start}+T'}_{T_{\rm start}} dt  \, 
 e^{i\omega t} {\bf J}^{\rm probe}(t)  F\left( \frac{t-T_{\rm start}}{T'}\right)
\end{equation}
where $T_{\rm start}=T_{\rm delay}-{T^{\rm probe}}/{2}$ and $T'=20$ fs. We have used a smoothing function $F(x)=1-3x^2+2x^3$.
The Fourier transform of the probe electric field is  defined in the same manner.

\section{Results and discussion \label{sec:results}}

\subsection{TDDFT results\label{sec:tddft_results}}

\begin{figure}
    \includegraphics[keepaspectratio,width=\columnwidth]{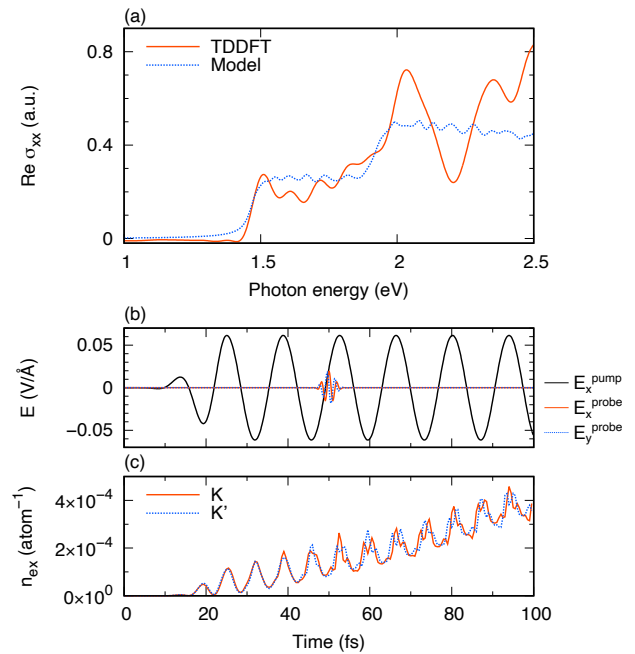}
    \caption{\label{fig:sigma} 
    (a) Real part of the $xx$ component of the conductivity of the WSe$_2$ monolayer calculated by TDDFT and the two-band model.
    (b) Time profiles of the pump and probe electric fields with $T_{\rm delay}=$ 50 fs.
    (c) Number of excited electrons near the respective valleys  caused by the pump field (TDDFT).
    }
\end{figure}
First, we discuss the  results obtained by TDDFT.
Figure~{\ref{fig:sigma}}(a) shows the real part of the conductivity, ${\rm Re}\,\sigma_{xx}(\omega)$, of the WSe$_2$ monolayer calculated by the linear-response TDDFT method in the absence of an external field (red-solid line).
For comparison, we  illustrate the results obtained by the two-band model (blue-dotted line) as well, which is scaled up by an arbitrary factor.
Figure~{\ref{fig:sigma}}(b) shows the time profile of the pump electric field (black line) defined in Eq.~(\ref{eq:pump}) and that of the probe electric field with $T_{\rm delay}=$ 50 fs (red and blue lines) defined in Eq.~(\ref{eq:probe}).
Figure~{\ref{fig:sigma}}(c) shows the TDDFT results for the number of excited electrons around the $K$ and $K'$ valleys [Eq.~(\ref{eq:nex_tddft})] in the presence of only the pump field. 
Although the number of excited electrons incrementally increases by small degrees owing to the multiphoton absorption, the amount of excitation is extremely low because the pump field is nonresonant. 
As expected for the pump field with the linear polarization, the deviation between the $K$ and $K'$ valleys is almost negligible.

Furthermore, we irradiate the system with the circularly-polarized probe pulse and calculate the pump-modulated transient conductivity using TDDFT.
\begin{figure}
    \includegraphics[keepaspectratio,width=\columnwidth]{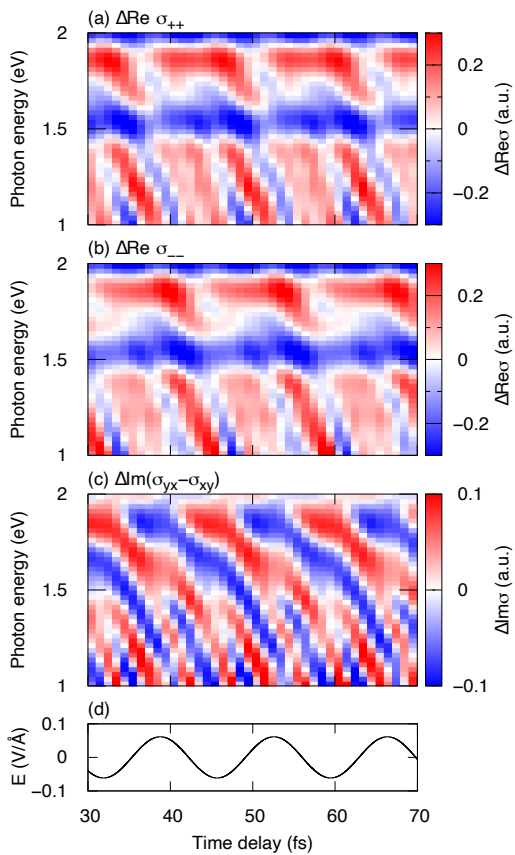}
    \caption{\label{fig:dsigma_tddft} 
    Transient conductivity change calculated via TDDFT as a function of the time delay $T_{\rm delay}$ and the  photon energy $\hbar \omega$ of the probe pulse.
    (a) Left-circular polarization result  ${\rm Re} \,\sigma_{++}(\omega,T_{\rm delay})$.
    (b) Right-circular polarization result  ${\rm Re} \,\sigma_{--}(\omega,T_{\rm delay})$.
    (c) Difference between the two ${\rm Im} \left[
     \sigma_{yx}(\omega,T_{\rm delay}) - \sigma_{xy}(\omega,T_{\rm delay})
    \right]$.
    (d) Applied pump field $E^{\rm pump}_x(t)$.
    }
\end{figure}
Figure~\ref{fig:dsigma_tddft} shows the transient conductivity change from the ground state value, calculated using TDDFT and via Eq.~(\ref{eq:sigma_circular_probe}), as a function of the time delay $T_{\rm delay}$ and the photon energy $\hbar \omega$ of the probe pulse.
Figure~\ref{fig:dsigma_tddft}(a) and (b) show the real part of the circular conductivity, 
${\rm Re} \,\sigma_{\pm\pm}(\omega,T_{\rm delay})$,
for the left and right circular probe pulses, respectively [refer to Eq.~(\ref{eq:work})].
Figure~\ref{fig:dsigma_tddft}(c) presents the difference between Fig.~\ref{fig:dsigma_tddft}(a) and (b) corresponding to the circular dichroism [refer to Eq.~(\ref{eq:work_diff})], and is given by the imaginary part of the off-diagonal conductivity ${\rm Im} \left[
     \sigma_{yx}(\omega,T_{\rm delay}) - \sigma_{xy}(\omega,T_{\rm delay})
    \right]$.
Figure~\ref{fig:dsigma_tddft}(d) shows the applied pump field $E^{\rm pump}_x(t)$ for comparison.
We note that the the bandgap is 1.5~eV and the time period of the pump field is $T_{\Omega}=2\pi/\Omega=$ 13.8 fs.
In Fig.~\ref{fig:dsigma_tddft}(a,b), the oscillation of the transient conductivity along the axis of the time delay behaves as a superposition of the $\Omega$ and $2\Omega$ oscillations.
The blue region around the bandgap indicates the peak reduction by pump-driven DFKE, whereas the red region below the bandgap implies the red-shift of the absorption edge.
In Fig.~\ref{fig:dsigma_tddft}(c), the off-diagonal transient conductivity exhibits the $\Omega$ ($3\Omega$) oscillation above (below) the bandgap.
Generally, the $n\Omega$ oscillation ($n$ is an odd number) of the transient conductivity in ATAS is a characteristic property of non-centrosymmetric systems because odd-order harmonics are forbidden in centrosymmetric systems \cite{Yamada2020}.
Although the the $\Omega$ oscillation phase near the bandgap correlates with the pump electric field phase, it gradually shifts as the photon energy moves farther from the bandgap.
As desired, the phase of the $\Omega$ oscillation is flipped depending on the left or right polarization of the probe pulse.
Thus, this phase-flipped oscillation is regarded as the valley-selective DFKE.

\begin{figure}
    \includegraphics[keepaspectratio,width=\columnwidth]{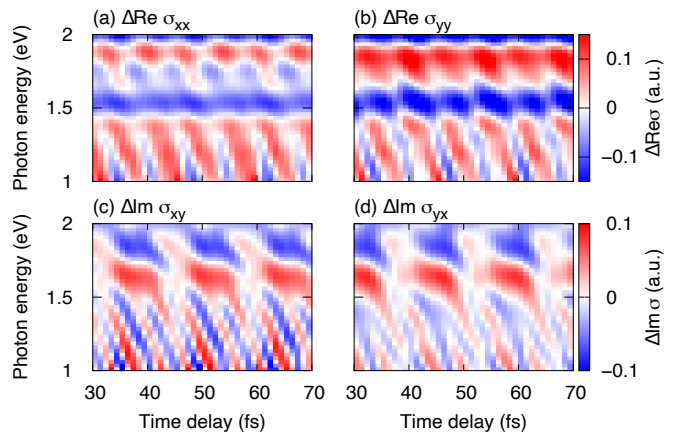}
    \caption{\label{fig:dsigma_tddft_tensor} 
    Tensor components of the conductivity change calculated via TDDFT.
    (a,b) Real  part  of the diagonal components.
    (c,d) Imaginary part  of the off-diagonal components.
    }
\end{figure}
Figure~\ref{fig:dsigma_tddft_tensor} shows the tensor components in the Cartesian coordinate calculated  by using Eq.~(\ref{eq:sigma_probe_tensor}) from the same data.
The upper panels (a) and (b) [lower panels (c) and (d)] are the real (imaginary) part  of the diagonal (off-diagonal) components.
While the oscillation of the diagonal components is mainly the $2\Omega$ oscillation, that of the off-diagonal components is mainly the $\Omega$  ($3\Omega$) oscillation above (below) the bandgap.
These observations are consistent with a symmetry consideration for DFKE with the reflection symmetry in the $x$ axis\cite{Yamada2020}.
From Eq.~(60) [Eq.~(61)] in Ref.~\onlinecite{Yamada2020}, the diagonal (off-diagonal) elements of the transient conductivity show oscillation with an even (odd) multiple of  the frequency $\Omega$ for the pump field polarized along the  reflection symmetry axis.
From Eq.~(\ref{eq:sigma_circular_probe}), we consider that the phase-flipped $\Omega$ oscillation of the circular components originates from the $\Omega$ oscillation of the off-diagonal components of the transient conductivity.

\begin{figure}
    \includegraphics[keepaspectratio,width=\columnwidth]{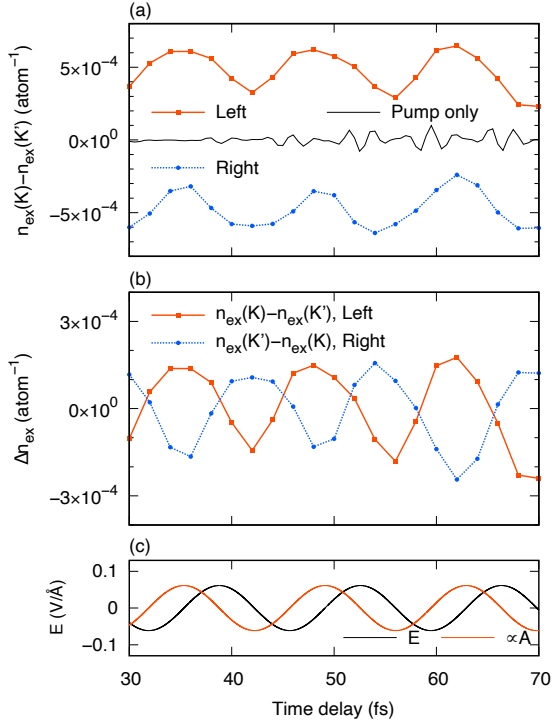}
    \caption{\label{fig:nex_tddft} 
    (a) Difference of the number of excited
    electrons between the $K$ and $K'$ valleys induced by the pump plus probe field [Eq.~(\ref{eq:nex_probe})].
    The red (blue) line corresponds to the case of the left (right) circular polarization.
    The black thin line is for the pump-only case.
    (b) The same as (a) but subtracting its mean value from each result. The sign is inverted for the right-circular result.
    (c) Applied pump field $E^{\rm pump}_x$ (black line) and corresponding vector potential (red line), where the latter is scaled up by an arbitrary factor.
    }
\end{figure}

Subsequently, we discuss the valley polarization of the excited electron population  by the pump plus probe field.
As discussed earlier, although we use the term ``pump-probe," both pulses act as pump fields for the excited electron population.
Figure~\ref{fig:nex_tddft}(a) shows the time-delay dependence of the following time-averaged value:
\begin{equation}
 \int_{T_{\rm end}}^{T_{\rm end}+T'} \frac{dt}{T'}  [n_{{\rm ex},K}^{\rm pump+probe}(t) - n_{{\rm ex},K'}^{\rm pump+probe}(t)],
 \label{eq:nex_probe}
\end{equation}
where $T_{\rm end}=T_{\rm delay}+{T^{\rm probe}}/{2}$ (end time of the probe pulse) and $T'=20$ fs.
Here, $n_{{\rm ex},K}^{\rm pump+probe}(t)$ [$n_{{\rm ex},K'}^{\rm pump+probe}(t)$] denotes the number of excited electrons near the $K$ ($K'$) point by the pump plus probe field calculated via Eq.~(\ref{eq:nex_tddft}).
This difference value means an imbalance, or valley polarization, of the electron excitation between the $K$ and $K'$ valleys. 
The red-solid (blue-dotted) line corresponds to that with the probe pulse of the left (right) circular polarization.
For comparison, the value for the pump-only case [deviation between the lines in Fig.~\ref{fig:sigma}(c)] is  plotted  by the black-thin line.
The left (red line) and right (blue line) circular cases exhibit $\Omega$ oscillation with respect to the mean values, whereas the pump-only case (black line) is almost zero.
The mean values of the red and blue lines are caused by the valley-selective excitation with the circular polarization.
The $\Omega$ oscillation of the red and blue lines originates from the variations in the transient absorption  by DFKE, as depicted in Fig.~\ref{fig:dsigma_tddft}(c).
The phase of the $\Omega$ oscillation corresponds to that of the vector potential of the pump field [red line in Fig.~\ref{fig:nex_tddft}(c)].

In Fig.~\ref{fig:nex_tddft}(b), to emphasize the oscillation behavior, we plot the same data but subtracted its mean value over the time delay axis from each result, where  the  sign for the right-circular result is inverted.
The red-solid (blue-dotted) line in Fig.~\ref{fig:nex_tddft}(b) describes the amount of change in the valley polarization induced by DFKE relative to the the $K$ ($K'$) valley, where the left (right) circular probe pulse is coupled to that valley.
As indicated in Fig.~\ref{fig:nex_tddft}(b),  the valley polarization change exhibits opposite phases depending on the helicity of the probe pulse.
Thus, the results in Fig.~\ref{fig:nex_tddft}(b) supports that the left (right) circular result corresponds to DFKE in the $K$ ($K'$) valley and
the valley-selective DFKE has been achieved by the pump-probe calculations.

\subsection{Model results}

\begin{figure}
    \includegraphics[keepaspectratio,width=\columnwidth]{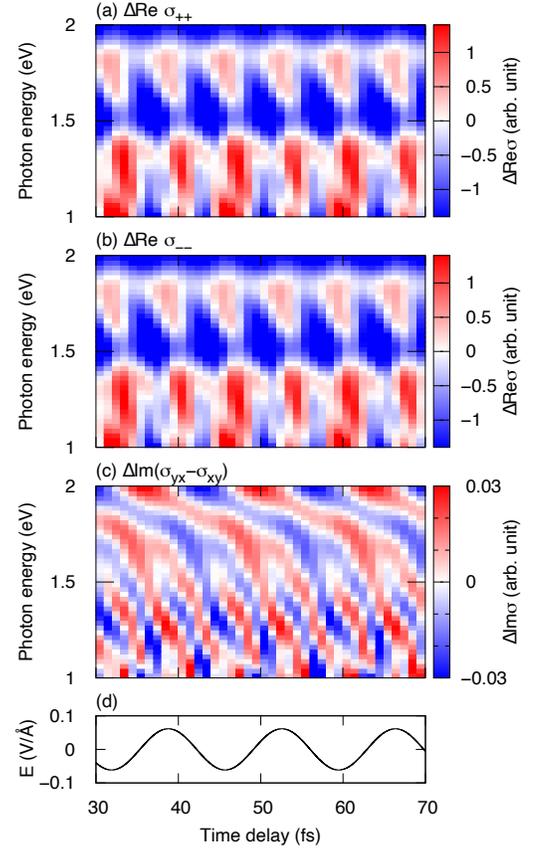}
    \caption{\label{fig:dsigma_model} 
    The same as Fig.~\ref{fig:dsigma_tddft} but for the two-band model.
    }
\end{figure}
To elucidate the key elements causing the valley-selective DFKE, we perform the same calculations as earlier but with the two-band model in Sec.~\ref{sec:model}.
Figure~\ref{fig:dsigma_model} shows the transient conductivity change and is the same as Fig.~\ref{fig:dsigma_tddft} but for the two-band model.
Unlike the TDDFT results, the transient conductivity in Fig.~\ref{fig:dsigma_model}(a,b) mainly exhibits  the $2\Omega$ oscillation but the $\Omega$ oscillation is very weak.
However, the off-diagonal conductivity in Fig.~\ref{fig:dsigma_model}(c) is qualitatively the same as that in Fig.~\ref{fig:dsigma_tddft}(c).
Except for the strong $2\Omega$ oscillation, the two-band model results qualitatively reproduce the TDDFT results.

\begin{figure}
    \includegraphics[keepaspectratio,width=\columnwidth]{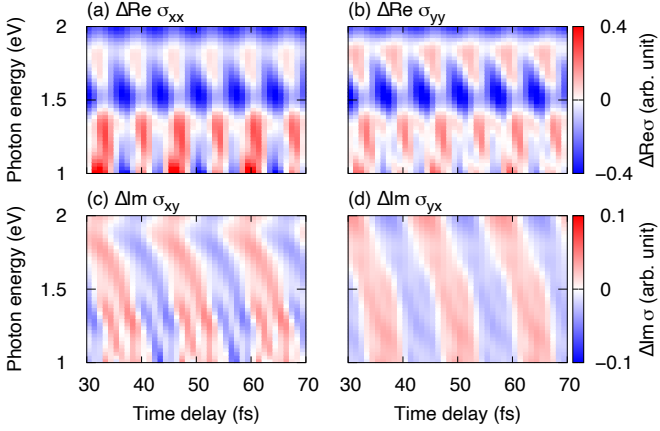}
    \caption{\label{fig:dsigma_model_tensor} 
    The same as Fig.~\ref{fig:dsigma_tddft_tensor} but for the two-band model.
    }
\end{figure}
Figure~\ref{fig:dsigma_model_tensor} is the same as Fig.~\ref{fig:dsigma_tddft_tensor} but for the two-band model.
As in the case of TDDFT, the oscillation of the diagonal (off-diagonal) components is primarily the $2\Omega$ oscillation ($\Omega$ oscillation).
Using the two-band model, we have successfully reproduced the valley-selective DFKE in the TDDFT case.

To identify the key elements for the valley-selective DFKE, we eliminate the second-order Hamiltonian [the second term in Eq.~(\ref{eq:model})] and re-evaluate the pump-probe calculations.
In this case, we confirmed that the $\Omega$ oscillation in the transient conductivity disappears and the off-diagonal elements, $\sigma_{xy}(\omega, T_{\rm delay})$ and $\sigma_{yx}(\omega, T_{\rm delay})$, are equal to zero. 
The diagonal elements contain only the $2\Omega$ oscillation.
Therefore, as  expected, the phase inversion of the $\Omega$ oscillations in the valley-selective DFKE is caused by the band asymmetry represented by the second-order Hamiltonian in the two-band model.

\begin{figure}
    \includegraphics[keepaspectratio,width=\columnwidth]{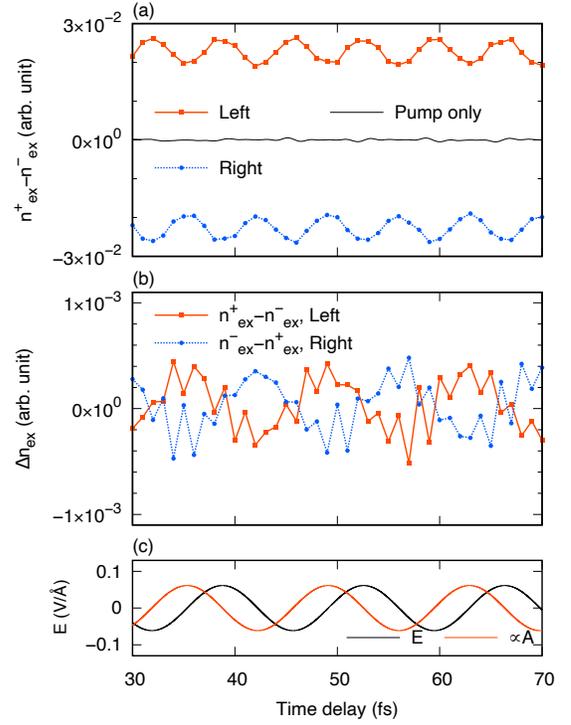}
    \caption{\label{fig:nex_model} 
    (a) The same as Fig.~\ref{fig:nex_tddft}(a) but for the two-band model.
    (b) Difference between the same data as (a) and the data without the second-order Hamiltonian. The sign of the right-circular result is inverted as Fig.~\ref{fig:nex_tddft}(b).
    (c) The same as Fig.~\ref{fig:nex_tddft}(c).
    }
\end{figure}

\begin{figure}
    \includegraphics[keepaspectratio,width=\columnwidth]{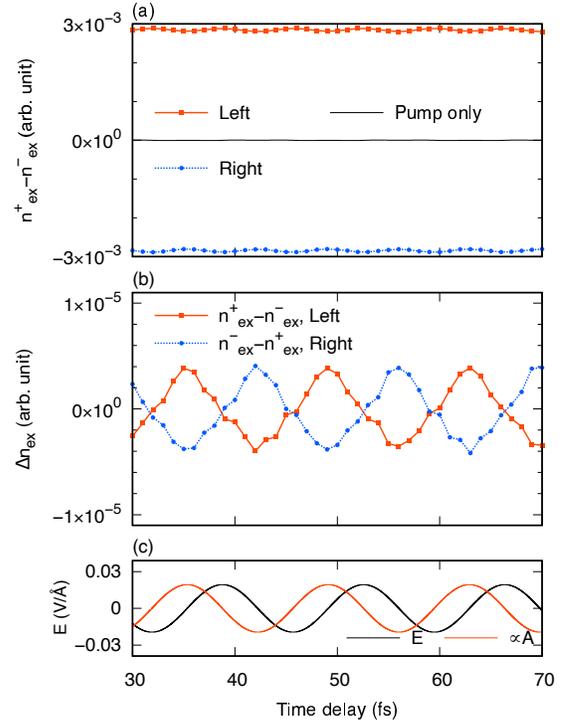}
    \caption{\label{fig:nex_model_weak} 
    The same as Fig.~\ref{fig:nex_model}, but with $I_{\rm pump}=5 \times 10^{9}$ W/cm$^2$.
    }
\end{figure}

Finally, we observe the number of excited electrons for the two-band model [Eq.~(\ref{eq:nex_model})].
Figure~\ref{fig:nex_model}(a) is the same as Fig.~\ref{fig:nex_tddft}(a) but for the two-band model.
Unlike the TDDFT results, the two-band model primarily demonstrates the $2\Omega$ oscillation.
As the $2\Omega$ oscillation is insignificant in the valley-selective DFKE, we shall extract the $\Omega$ oscillation.
The deviations between the  data in Fig.~\ref{fig:nex_model}(a) and that obtained without the second-order Hamiltonian are depicted in Fig.~\ref{fig:nex_model}(b), where the sign of the right-circular result is inverted similar to that in Fig.~\ref{fig:nex_tddft}(b).
Although the oscillation behavior in Fig.~\ref{fig:nex_model}(b) is dirty due to higher-order nonlinear terms, the $\Omega$ oscillation is qualitatively similar to that in the TDDFT case of Fig.~\ref{fig:nex_tddft}(b).
Similar to Fig.~\ref{fig:nex_model}, the results for a weaker pump field of $I_{\rm pump}=5 \times 10^{9}$ W/cm$^2$ are illustrated in Fig.~\ref{fig:nex_model_weak}.
In Fig.~\ref{fig:nex_model_weak}(b), the higher-order oscillations are suppressed and the similarity of the $\Omega$ oscillation to the TDDFT case [Fig.~\ref{fig:nex_tddft}(b)] becomes apparent. 


As discussed earlier, the limitation of the two-band model is exposed by Fig.~\ref{fig:dsigma_model}(a,b) and Fig.~\ref{fig:nex_model}, where the higher-order oscillations are apparent compared to the TDDFT case.
Originally, the two-band model Eq.~(\ref{eq:model}) has aimed  to describe the low energy physics around the $K$ or $K'$ valley.
Unsurprisingly, the two-band model fails to quantitatively reproduce the nonlinear excitation in the first-principles calculations because the model cannot represent the effects of multiple bands nor the entire Brillouin zone.
Nevertheless, the two-band model can reproduce the qualitative behavior of the TDDFT results and reveal the physical mechanisms governing the valley-selective DFKE.

\section{Conclusions \label{sec:conclusion}}

In summary, this study performed first-principles calculations based on TDDFT to ensure the feasibility of pump-probe measurements for the valley-selective DFKE in TMDC monolayers.
The numerical pump-probe experiments for the WSe$_2$ monolayer demonstrated that the valley-selective DFKE oscillation in the transient conductivity was achieved by the linearly-polarized pump field and the circularly-polarized probe pulse.
The results revealed that the transient conductivity and excitation probability around each valley oscillate with the same frequency $\Omega$ as that of the pump field, and the phases of the oscillation for the $K$ and $K'$ valleys are opposite to each other.
The helicity of the probe pulse corresponds to the valley selection and controls the phase flipping of the DFKE oscillation.
For the transient conductivity, the phase-inverted  oscillation of the circular components is caused by the odd-order oscillations in the off-diagonal components.
Moreover, the valley polarization of the excited electron population exhibits the phase-inverted $\Omega$ oscillation owing to the transient absorbance change induced by the valley-selective DFKE.
Using the simplified two-band model, we qualitatively  reproduced the TDDFT results  and identified the band asymmetry around the $K$ or $K'$ valley as the key element causing the valley-selective DFKE.
Thus, the present findings established the valley-selective optical switching using DFKE in TMDC monolayers for future research and development in this area.

\begin{acknowledgements}
This research was supported by JST-CREST under grant number JP-MJCR16N5, and by MEXT Quantum Leap Flagship Program (MEXT Q-LEAP) Grant Numbers JPMXS0118068681 and JPMXS0118067246, and by JSPS KAKENHI Grant Number 20H2649. 
Calculations were performed on Fugaku supercomputer with support from the HPCI System Research Project (Project ID: hp220120), SGI8600 at Japan Atomic Energy Agency (JAEA), and Wisteria at the University of Tokyo under  Multidisciplinary Cooperative Research Program in CCS, University of Tsukuba.
\end{acknowledgements}

\input{reference.bbl}

\end{document}

%% file: reference.bbl
%